# Alpha Particle X-Ray Spectrometer (APXS) On-board Chandrayaan-2 Rover - Pragyan

M. Shanmugam[1], S. V. Vadawale[1], Arpit R. Patel[1], N. P. S. Mithun[1], Hitesh kumar Adalaja[1], Tinkal Ladiya[1], Shiv Kumar Goyal[1], Neeraj K. Tiwari[1], Nishant Singh[1], Sushil Kumar[1], Deepak Kumar Painkra[1], A. K. Hait[2], A. Patinge[2], Abhishek Kumar[3], Saleem Basha[3], Vivek R. Subramanian[3], R. G. Venkatesh[3], D. B. Prashant[3], Sonal Navle[3], Y. B. Acharya[1]  S. V. S. Murty[1], and Anil Bhardwaj[1]

[1]Physical Research Laboratory, Ahmedabad
[2]Space Application Centre, Ahmedabad
[3]U. R. Rao Satellite Centre, Bengaluru
*Email: shansm@prl.res.in*

## Abstract

Alpha Particle X-ray Spectrometer (APXS) is one of the two scientific experiments on Chandrayaan-2 rover named as Pragyan. The primary scientific objective of APXS is to determine the elemental composition of the lunar surface in the surrounding regions of the landing site. This will be achieved by employing the technique of X-ray fluorescence spectroscopy using in-situ excitation source $^{244}$Cm  emitting both X-rays and alpha particles. These radiations excite  characteristic X-rays of the elements by the processes of particle induced X-ray emission (PIXE) and X-ray fluorescence (XRF). The characteristic X-rays are detected by the 'state-of-the-art' X-ray detector known as Silicon Drift Detector (SDD), which provides high energy resolution as well as high efficiency in the energy range of 1 to 25 keV. This enables APXS to detect all major rock forming elements such as, Na, Mg, Al, Si, Ca, Ti and Fe. The Flight Model (FM) of the APXS payload has been completed and tested for various instrument parameters. The APXS provides energy resolution of ~135 eV at 5.9 keV for the detector operating temperature of about -35°C. The design details and the performance measurement of APXS are presented in this paper.

**Introduction**

Chandrayaan-2, the second Indian mission to the Moon has a soft lander, named Vikram, which contains the Pragyan rover. The rover, provides a unique opportunity to conduct the in-situ measurement of elemental composition in the regions near the landing site at unprecedented details. Quantitative estimates of elemental abundances on local scales play an important role in understanding the evolution of the lunar surface[1]. These measurements are very important, and hence the Chandrayaan-2 rover carries two experiments to determine the elemental composition using two independent techniques – (i) LIBS (Laser Induced Breakdown Spectrometer)[2], which measures elemental concentration based on the optical lines from the laser induced plasma emission; and (ii) APXS (Alpha Particle X-ray Spectrometer), which measures elemental concentration based on the characteristic X-ray lines emitted by the atoms excited by incident alpha and X-ray radiation. The technique of determining elemental abundances using characteristic X-ray lines through APXS instrument is well proven by a series of experiments on Martian surface since the three generations of Mars rovers, namely - Sojourner rover on the Mars Pathfinder mission in 1997[3], Spirit and Opportunity (Mars Exploration Rovers) in 2003[4], and Curiosity (Mars Science Laboratory) in 2013[5], have provided a wealth of data on the elemental composition at various locations of the Martian surface[6-8]. The APXS instrument was also carried on-board the Philae lander of the Rosetta mission to study the comet 67P/CG[9], but unfortunately it could not be utilized due to the issues with landing of the lander. In the context of lunar studies, in-situ measurements on the Moon were initiated in the late 1960s through detection of back scattered alpha particles with Surveyor lander program on Moon[10]. However, after the Apollo missions in early 1970s, there was a long break in the lunar landing missions with next lunar landing taking place after almost four decades in December 2013 by the Chinese mission Chang'e-3[11] and its follow-up landing mission Chang'e-4[12] in January 2019. The YUTU rover of Chang'e-3 had an XRF experiment to measure elemental composition[13]; however, the YUTU-2 rover of Chang'e-4 does not have any experiment to determine the elemental composition.

The Chandrayaan-2 Vikram landing site is planned to be in the lunar highland region close to the lunar South Pole. So far, all lunar landings, including the historic Apollo/Luna missions, as well as recent Chang'e-3 mission, have occurred in the mare regions. Thus, APXS will provide the in-situ measurements of elemental composition in the highland regions of the Moon for the first time in the history of the lunar exploration. Another important feature of APXS is the use of alpha particles as excitation source, which provide enhanced sensitivity for the lower atomic number elements, such as Na, Mg, Al, Si and Ca. It may be noted that the XRF experiment on-board the Chang'e-3 rover had only X-ray source as excitation radiation and thus had low sensitivity for low-Z elements. The APXS on-board Chandrayaan-2 rover is likely to provide a very unique data on the elemental composition in the lunar polar regions.

The APXS experiment for Chandrayaan-2 rover was finalized in 2010 and its design and development started thereafter. The initial version of APXS instrument design is presented in Shanmugam et al. (2014) [14]. However, the APXS has substantially evolved. Here we present the design details of the flight model of APXS instrument. The next sections outlines few salient features of APXS, followed by the instrument description, its performance and calibration as well as a brief discussion on the operation plan and data analysis.

## Scientific Objective

The primary scientific objective of APXS is to determine the elemental composition of the lunar surface in the regions surrounding the landing site. This will be achieved by employing the technique of X-ray fluorescence spectroscopy using in-situ excitation source. For this purpose, APXS uses $^{244}$Cm radio-active sources emitting both alpha particles having energy of 5.8 MeV and X-rays having energy of 14.3 and 18.3 keV, which excites the elemental characteristic X-rays by the processes of particle induced X-ray emission (PIXE) and X-ray fluorescence (XRF). The characteristic X-rays are detected by the 'state-of-the-art' X-ray detector known as Silicon Drift Detector (SDD), which provides high energy resolution as well as high detection efficiency in the energy range from 1 to 25 keV. This enables APXS to detect all major rock forming elements, such as Na, Mg, Al, Si, Ca, Ti, Fe, etc. and some trace elements, such as Sr, Y, Zr which are found in the lunar regolith. This is the first time such measurements will be carried out on the Moon.

## APXS salient features

**Cm-244 radio-active source**

One of the key factors enabling APXS to make sensitive measurements is the alpha source $^{244}$Cm, having half life of 18.1 years and emitting alpha particles with energy of 5.8 MeV and prominent X-ray lines with energies of 14.3 and 18.3 keV. The alpha particles enhance the sensitivity to low Z elements detection by more than two orders of magnitude, as these elements are primarily excited by alpha particles due to their low fluorescence yield. The $^{244}$Cm source has been used earlier in APXS experiments on Mars rovers as well as on Philae lander on Comet 67-P, however, this will be the first use of $^{244}$Cm in a lunar APXS experiment.

**Energy range of the spectrometer**

APXS is designed to measure the X-rays in the energy range 1 – 25 keV. The lower energy threshold of APXS is ~800 eV, which enables unambiguous detection of all elements beyond Na (Z=11, Kα line energy of 1.04 keV). For most of the major elements up to Z=30, the Kα line energy is less than 10 keV and for elements with Z > 30, the L-lines are within the energy range of 1 – 15 keV. However, the higher energy range of APXS has been intentionally kept at 25 keV, so as to cover the back scattered X-ray lines from the $^{244}$Cm source itself. These scattered lines are likely to have some signature of the very low-Z elements (X-ray invisible) due to Rayleigh and Compton scattering processes. Thus the high energy range of 25 keV enables the possibility of estimating low-Z elements abundance, particularly Oxygen, from the observed back scattered lines.

**Silicon Drift Detector**

In order to cover the required energy range $1 - 25$ keV with high resolution, APXS employs state-of-the-art Silicon Drift Detectors. Although SDDs are functionally similar to other thick Silicon based detectors, like Si-PIN or Si-Strip detectors, the specific electrode structure provides low detector capacitance resulting in better energy resolution in comparison with other detectors. SDD has been used in a similar experiment on-board Mars Science Laboratory (MSL)[5].

**APXS deployment mechanism**

The fluorescence signal detected by APXS strongly depends on the distance of sensor head from the target surface. From the experience of experiments with earlier versions of the APXS instrument, as well as simulations, it was determined that the exposure time required to obtain spectrum with sufficient statistics was about 5 to 6 hours[15]. Hence, in the new design of the APXS instrument, it is mounted on a deployment mechanism which brings the radio-active sources and the X-ray detector close to the lunar surface during observations and stowed back during rover movements to provide sufficient ground clearance. In the deployed condition, the X-ray detector will be about 55 mm above the lunar surface resulting in the circular field of view of about 12 cm diameter on the surface. This will yield measurement of fluorescence spectra with better accuracy in shorter exposure time.

**In-flight calibration**

A calibration target plate with multiple metallic sheets is mounted at the bottom of the rover chassis and covering the front face of the APXS in stowed position, which provides in-flight calibration as well as protection from dust particles during rover movements. The fluorescence line energies from the calibration target materials allow to monitor the gain and offset of the instrument. Any variation in the line intensities will allow us to determine the possible lunar dust contamination on the source or detector. Any deviations from ground calibration of the instrument, if present, can be quantified with the calibration plate observations and these corrections can be applied in the analysis of the in-flight data. As the calibration plate covers the sensor head of the APXS during rover movements, it would also curtail the possibility of levitated dust to deposit on the source or detector that could cause reduction of incident alpha particle energy and attenuation of low energy X-rays.

## APXS instrument description

The APXS payload is configured as a single package consisting of SDD, two Printed Circuit Boards (PCBs) mounted on the base of the package accommodating the readout and control electronics and a source holder assembly. Table 1 provides an overview of the APXS instrument specifications.

| Parameter | Value |
| --- | --- |
| Energy range | 1 – 25 keV |
| Detector | Silicon Drift Detector (SDD) |
| Detector size | 30mm$^2$, 450 micron thick |
| Cooling | In-built thermo electric cooler with SDD |
| Detector operating temperature | ~ -35ºC using inbuilt peltier |
| Resolution | ~135 eV at 5.9 keV |
| Source | $^{244}$Cm alpha source |
| Number of sources and total activity | 6 sources, each with ~5mCi activity (total: ~30 mCi) |
| Calibration target | 4 metal targets (Al, Ti, SS and Cu) |
| Quantization | 12 bits |

**Table 1: APXS instrument specifications**

The block schematic of APXS instrument and its interface with rover is shown in figure 1. A common rover Field Programmable Gate Array (FPGA) based electronics is responsible for the APXS payload operation and payload data acquisition as well as the mechanism operations.

The SDD detector and the source holder assembly are mounted at one end of the package and the other end of the package has a provision for interface with the motor based mechanism as shown in figure 2. The photographic view of the Flight Model (FM) of the APXS instrument is shown in figure 3.

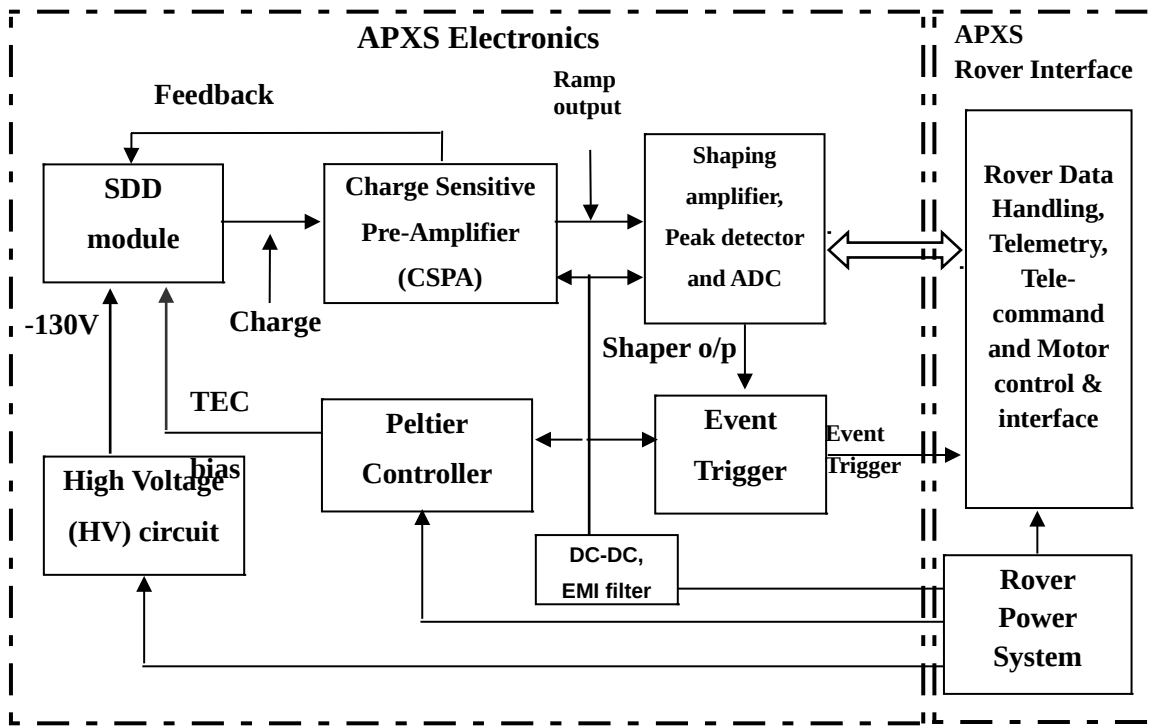

**Figure 1: Block schematic of APXS electronics and Pragyan rover interface.**

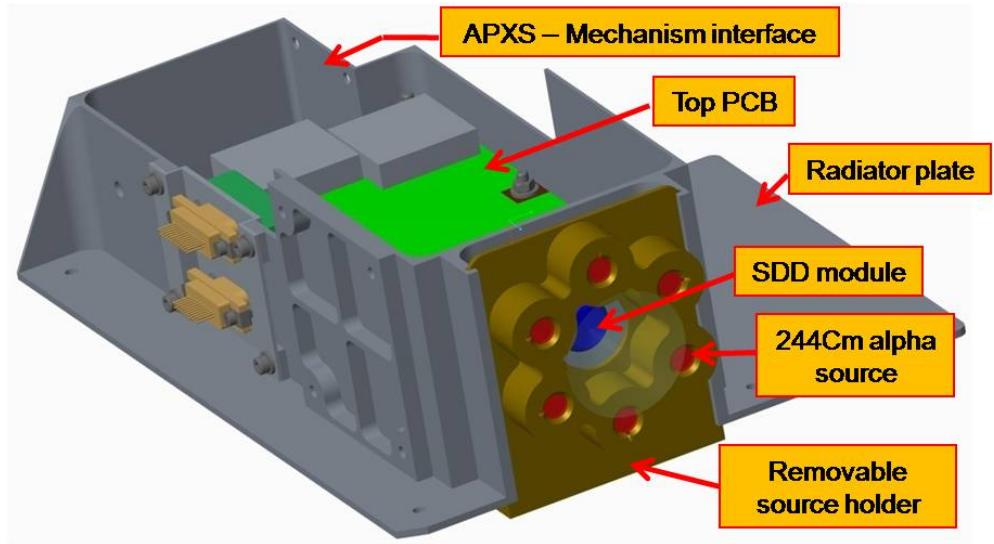

Figure 2: APXS mechanical model.

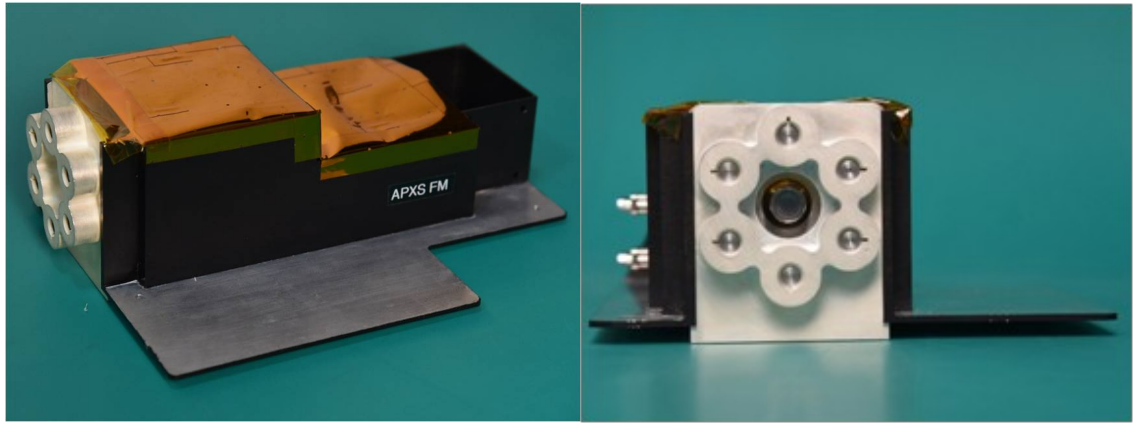

Figure 3: Flight model package of APXS - Side view (left) and front view (right).

**Detector and front-end electronics**

*SDD module*

SDDs are typically available in the form of modules containing an SDD chip interfaced with a Field Effect Transistor (FET), feedback capacitor and a reset diode. This forms the first stage of the charge readout electronics which is critical for the instrument performance. The SDD module used in the APXS experiment is procured from KETEK, GmbH because of its space heritage. The active area and thickness of the SDD are ~30 mm$^2$ and 450 µm respectively. SDD module also contains peltier cooler and a temperature sensor for precise temperature control of SDD chip operating temperature. The in-built peltier element has the capacity to achieve the maximum differential temperature of 75°C. The KETEK SDD module is available as a standard TO-8 package with an 8 µm thick Beryllium window on the top. Figure 4 shows the photographic view of the detector module with and without encapsulation.

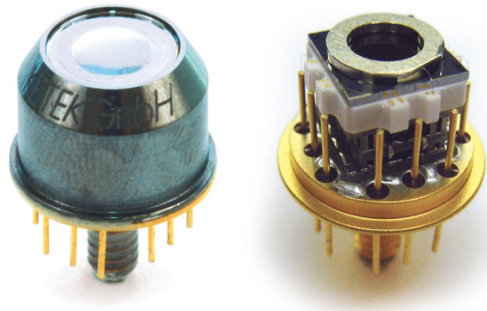

**Figure 4 : Photographic view of SDD module with and without external cover (Courtesy KETEK, GmbH, Germany).**

To achieve the desired energy resolution, the temperature of the SDD is maintained at around -35°C using closed-loop control of the thermo-electric cooler. The SDD detector module is mounted on an Aluminium wall with a radiator for efficient thermal management which is part of APXS package as shown in figure 2. The electrical interface with the SDD module and the front-end readout electronics is established through a flexi-rigid interface.

*Readout electronics*

APXS readout electronics consists of Charge Sensitive Pre-Amplifier (CSPA) that directly interfaces with SDD module, followed by a shaping amplifier with a baseline restorer (BLR). CSPA converts the charge cloud generated in the SDD due to incident X-ray in to voltage form. The "Reset type" CSPA design is adapted in APXS. The output of CSPA is fed to a shaping amplifier which converts the small step pulses in to semi-Gaussian pulses with required amplification. Shaping amplifier is a $CR-(RC)^2$ type three stage amplifier, which provides the pulse peaking time of ~ 2 μs with gain of ~25 for covering the energy range 1– 25 keV. Shaping amplifier includes a BLR that minimizes the baseline fluctuations for improved energy resolution. The output of shaping amplifier is then provided to an event trigger generator and a peak detector. The event trigger generator provides indication of X-ray interaction with the SDD when the shaping amplifier output crosses a set threshold value. On occurrence of the event trigger, rover FPGA generates the control signals for the peak detector to hold the peak signal amplitude for Analog to Digital Conversion (ADC). APXS uses a 12 bit serial Analog to Digital Converter. The rover FPGA reads 12 bit digital data and generates packets of ~1 KB containing individual X-ray events along with the header data. The readout electronics also includes the high voltage bias generation circuits for the SDD operation. A voltage multiplier based circuit provides the high voltage biases of -20 V, -130 V and -60 V, that are applied to inner ring $R_1$, outer ring $R_x$ and back contact $R_b$ of the SDD, respectively.

*Peltier Controller*

For stable performance of the instrument, it is necessary to maintain the SDD temperature at around -35°C. This is carried out by providing sufficient power to the inbuilt thermo-electric cooler and draining the heat by coupling the detector to the metallic portion of the payload assembly along with the radiator plate. APXS readout electronics also incorporates an active peltier controller circuit which maintains the SDD operating temperature at about -35°C for the ambient temperatures less than +30°C. This is realized by controlling the gate voltage of N-Channel FET and thereby the peltier current, with respect to the difference between the set point and the measured temperature by the temperature sensor. It takes about two minutes for the peltier controller to achieve a stable nominal operating temperature from ambient conditions. Peltier controller has a provision to monitor the peltier current and detector temperature through the telemetry system.

*$^{244}$Cm source holder*

APXS carries six $^{244}$Cm alpha sources and these sources are housed in a source holder (see figure 2). The source holder is designed such that it can be assembled and removed from the instrument package whenever required considering the radiation safety aspects. The $^{244}$Cm alpha sources for the APXS experiment are procured from Russia. These sources were prepared by high temperature condensation of metal curium vapour onto a silicon disk substrates, forming silicide on the surface[16]. These discs are further encapsulated within the Titanium (Ti) alloy capsule and covered with 3 micron thick Ti foil on the front side to form a sealed source. Figure 5 shows the single encapsulated source capsule. The active diameter of the source is 6 mm, disc diameter is 8 mm and the encapsulated source diameter is 10 mm.

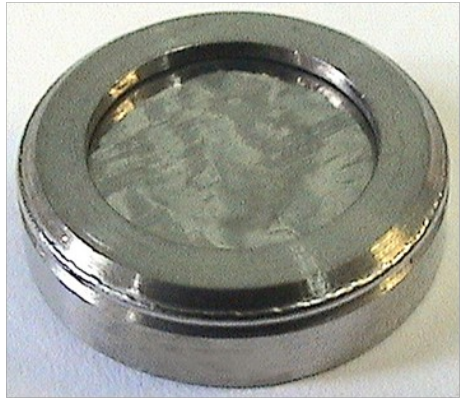

**Figure 5 : Encapsulated $^{244}$Cm alpha source.**

Spectral characterization of $^{244}$Cm sources revealed the presence of ~0.06 % $^{241}$Am, by activity, having a strong line at ~60 keV. Also, apart from the low energy lines, $^{244}$Cm nuclide has a line at around 44 keV. Although, the efficiency of 450 µm thick SDD is low at these high energies, given the large source activity, a fraction of these photons that are incident on the detector directly from source would contribute significantly to the observed spectrum. Photons depositing complete energy by photoelectric absorption are recorded in the last channel of the instrument (saturated events), which does not affect the observations except increasing the dead time slightly. However, a fraction of photons undergo Compton scattering in the detector depositing partial energies to produce a continuum spectrum in the low energy regime. This continuum at low energies affect the sensitivity of the instrument. Hence, it is required to block the high energy photons of the source from reaching the detector directly, by means of appropriate shielding around the source except the front side. This is accomplished by providing the graded Z shielding around the sources, considering the X-ray line energies and the edge energies of the shielding materials. A two mm thick gold cup encapsulating the source provides attenuation better than $1 \times 10^{-7}$ at 60 keV. Fluorescence lines arising from L transitions in the gold layer are absorbed by a 1 mm outer layer of stainless steel and the surrounding layer of aluminium that forms the source holder and absorbs secondary emission from the stainless steel. A last layer of 20 µm thick silver on the aluminium holder blocks the Al fluorescence lines. Figure 6 gives the exploded view of the source holder showing the gold and stainless steel encapsulations. This design has resulted in the reduction of continuum by about one order of magnitude as shown in the later section.

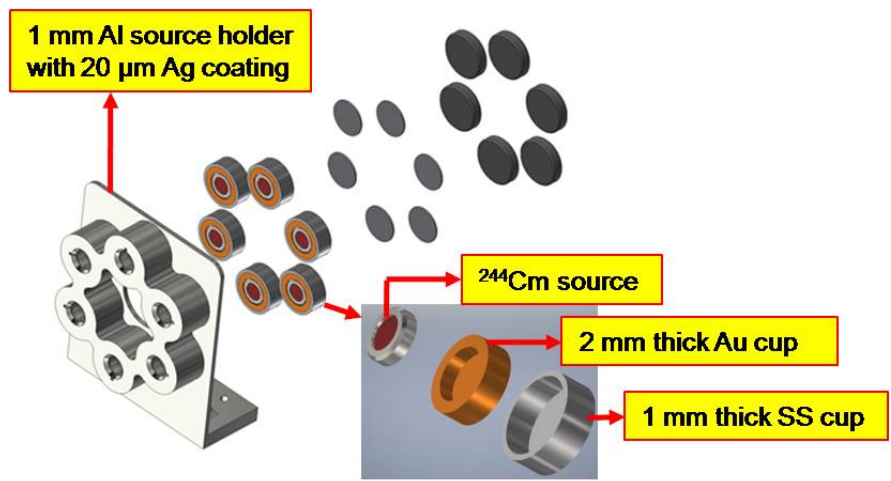

**Figure 6: APXS source holder assembly with successive layers of gold, stainless steel and the aluminium source holder coated with silver.**

*Deployment mechanism*

As discussed in the previous section, APXS employs a deployment mechanism which brings the instrument close to the lunar surface (~55 mm) during the measurements and is stowed back after the observation. APXS package is interfaced with the mechanism using an L clamp mounted on the shaft of brushless DC motor. This mechanism assembly is mounted on the bottom chassis of the rover as shown in figure 7. The mechanism steers the APXS from the stowed position (0º) to the deployed position (90º) and vice versa, and the micro-switches fixed at the extreme positions provide the feedback for controlling the deployment of the motor mechanism operation. In order to minimize the stress on the motor shaft during the launch and landing shocks, an additional support is provided to the instrument with the rover hold down mechanism, which is part of the rover. During the roll out of rover from lander ramp, this hold down will be released allowing the mechanism to steer APXS.

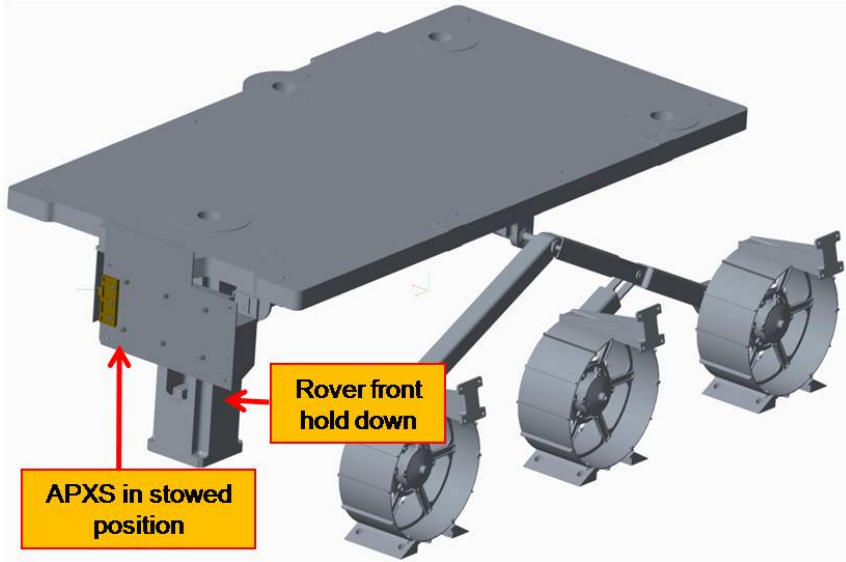

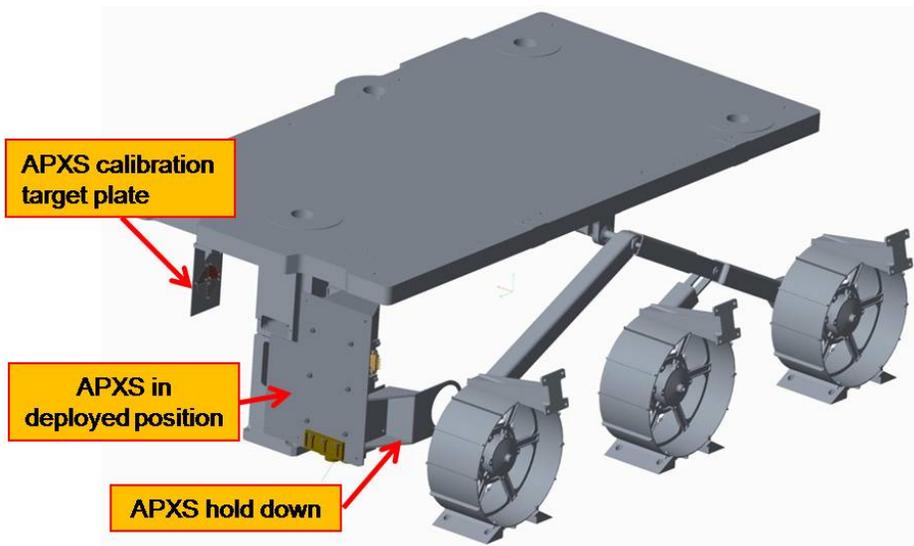

**Figure 7: APXS in stowed condition (top), deployed condition (bottom).**

*Design of Calibration target plate*

APXS carries a calibration target plate which is mounted on the bottom chassis of the rover, very close to the detector and the source assembly. The calibration target plate protects the detector and the sources from lunar dust contamination and also carries metallic targets for instrument calibration. The calibration target plate is made of aluminium and has four metallic targets, namely SS, Al, Ti and Cu as shown in figure 8. The target materials are selected such that they cover wide energy range for the fluorescent X-rays and also are sensitive to PIXE and XRF processes. For example, aluminium is excited mostly by Alpha particles whereas the copper target is excited by X-rays. This allows us to characterize any variations in alpha flux, alpha energy and X-ray flux from the source with time. The calibration target plate is also coated with 50 micron thick silver and 20 micron thick gold to provide additional radiation shielding when the actual source holder is integrated with the APXS on the rover.

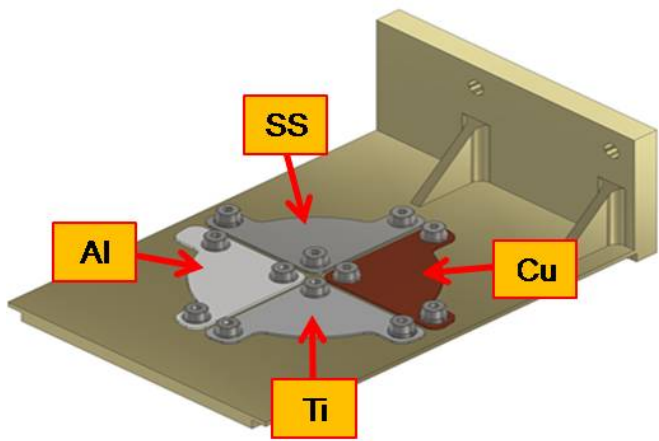

**Figure 8: Schematic representation of 4 metallic calibration targets used on-board APXS.**

## Instrument performance

The flight model of APXS is realized after verifying the performance of the qualification model during various environmental tests. As the data acquisition and control of APXS is carried out by a common rover FPGA, for the purpose of characterization at payload level, a ground checkout system is developed. This ground check out system is functionally similar to the FPGA based control and data acquisition system on the rover and generates data packets which are acquired on a computer using Labview-based data acquisition and control software. The test and evaluation as well as the detailed scientific calibration of the APXS instrument is carried out using this ground checkout system. To assess the performance of the instrument during the environmental tests, a $^{55}$Fe calibration source is used which is covered with titanium foil, having lines at 5.9, 6.5, 4.5 and 4.9 keV. Figure 9 shows the calibration source spectrum acquired for the nominal detector operating temperature of about -35°C. The spectral resolution measured as the full width at half maximum (FWHM) of the line at 5.9 keV is ~137 eV, which is better than the desired specification of 150 eV. The instrument performance after the environmental tests are same as that of the pre-test results.

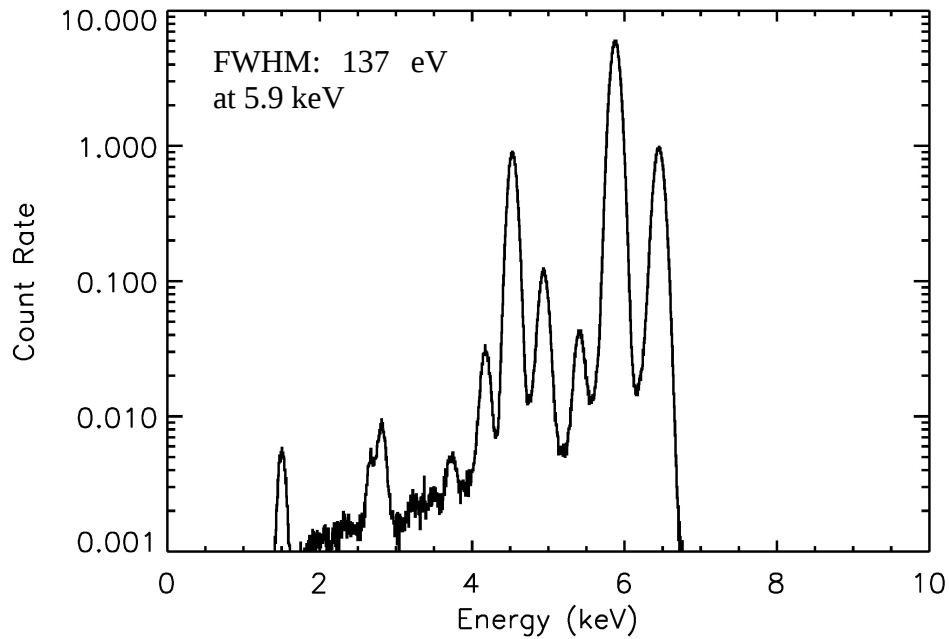

**Figure 9: APXS spectrum acquired with $^{55}$Fe X-ray source. Lines at 5.9, 6.5, 4.5 and 4.9 keV are prominent along with other fainter lines in the spectrum. The spectral resolution measured as FWHM of line at 5.9 keV is 137 eV.**

Further, to characterize the instrument along with $^{244}$Cm radioactive source assembly, experiments were carried out with different target materials, where the alpha and X-ray radiation from the source excites the elements and the fluorescence spectra are recorded. As alpha radiation energy and the low energy X-ray fluorescence line intensity attenuates while passing through the air, these experiments cannot be carried out at atmospheric pressure. Hence, to simulate the conditions of the Moon, the calibration experiment is carried out in the vacuum. Figure 10 shows the photograph of the experiment setup, where APXS is mounted in a vacuum chamber and has the provision to vary the target distances. Data is acquired for various metallic samples placed at different distances from APXS and figure 11 shows a sample spectrum for SS metal target placed at the nominal target distance of ~55 mm, where the fluorescence lines of the constituent elements are clearly identified.

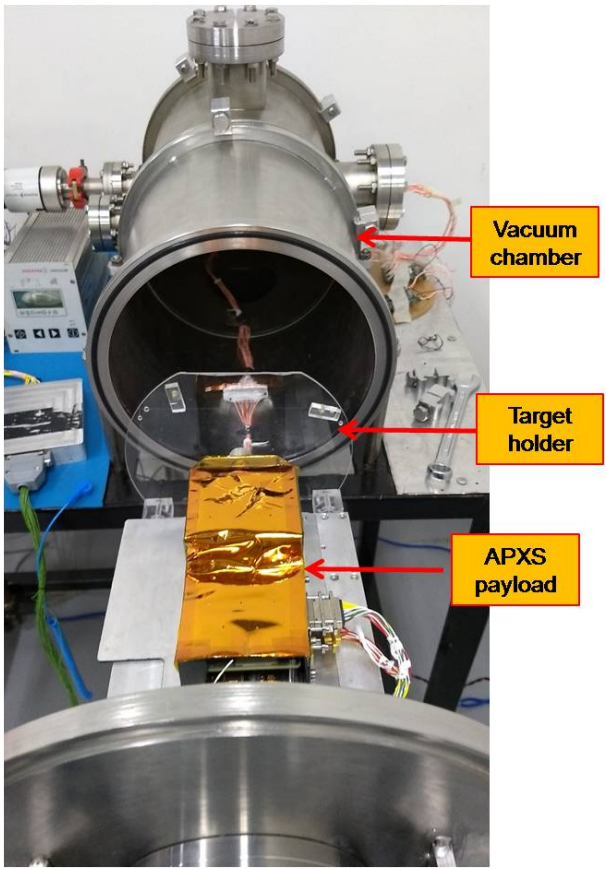

**Figure 10: Experimental setup for characterization of APXS with metallic targets in vacuum chamber. The metallic sample is mounted on a holder made of perspex. Distance of target from APXS can be varied by motorized movement of the target holder inside chamber.**

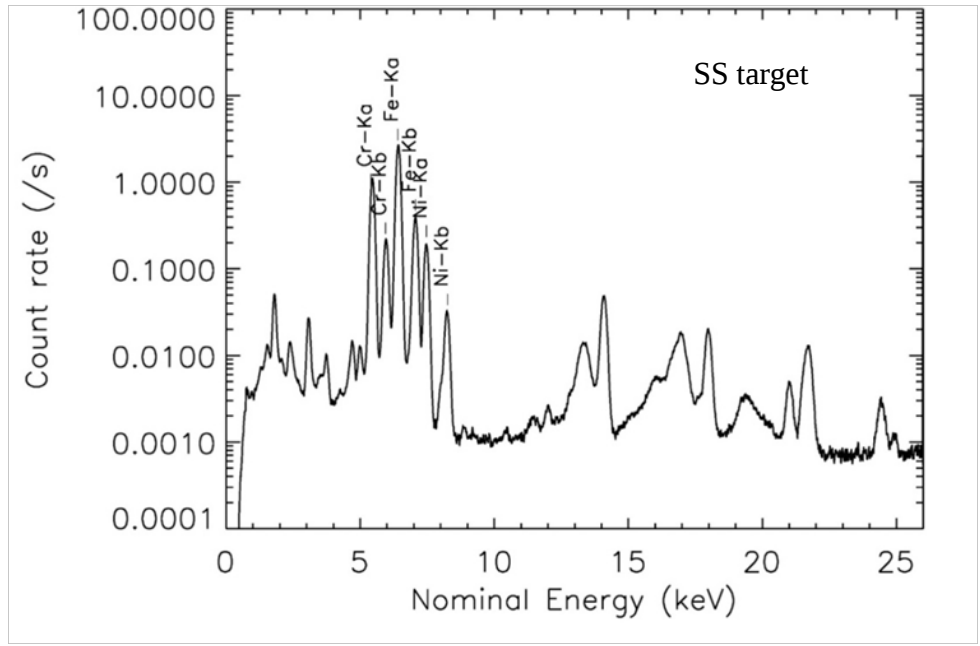

**Figure 11 : Spectrum acquired with APXS for stainless steel target.**

In order to derive the elemental composition of the unknown target from APXS spectrum, correlations between elemental line fluxes and compositions are to be derived. For this purpose, extensive experiments were carried out with geochemical samples with certified compositions and several other samples of unknown compositions, results of which will be reported later. One such spectra of a granite sample is shown in figure 12, along with spectra of the same sample blocking the alpha radiation from the source of APXS using a thin (250 µm) plastic sheet. It can be noted that in the latter case, lines of low Z elements are significantly weaker compared to high Z elements. This is expected as the low Z elements are primarily excited by PIXE, whereas the high Z elements by XRF and this demonstrates the requirement of alpha source for better sensitivity for low Z elements.

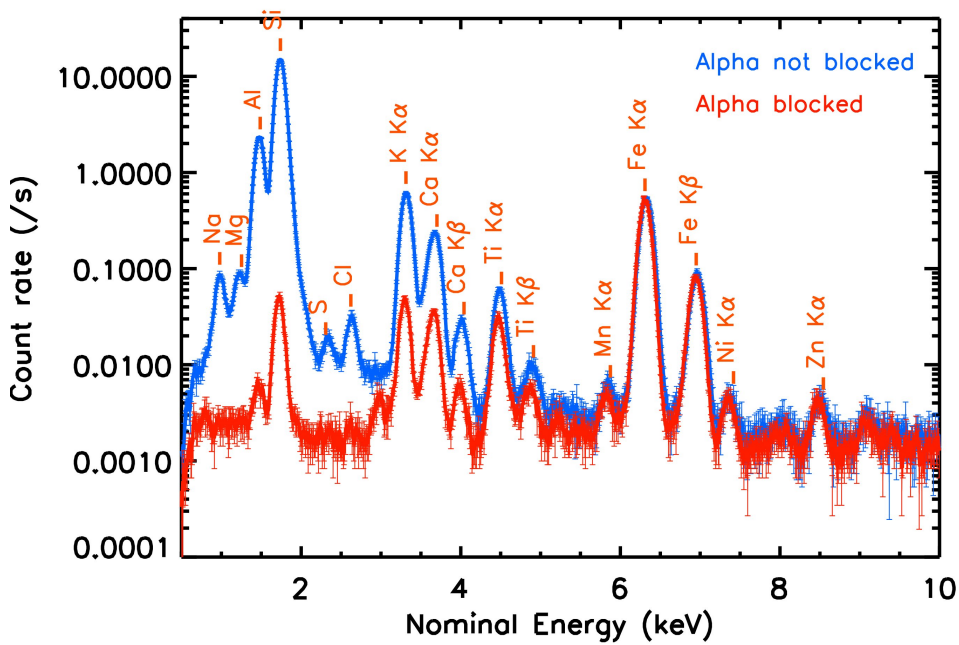

**Figure 12: APXS spectra of granite sample and spectra acquired with alpha radiation from source blocked using a thin plastic sheet.**

Another important feature of APXS which improves the sensitivity is the multilayer shielding of the $^{244}$Cm sources which blocks the high energy X-ray photons from directly reaching the detector as described earlier. Figure 13 shows the comparison of spectra acquired with qualification model of APXS incorporating the multilayer shielding with that of engineering model that does not include multilayer shielding, using perspex as target. It can be clearly seen that the continuum reduces by about one order of magnitude and there by improving the sensitivity for weaker lines.

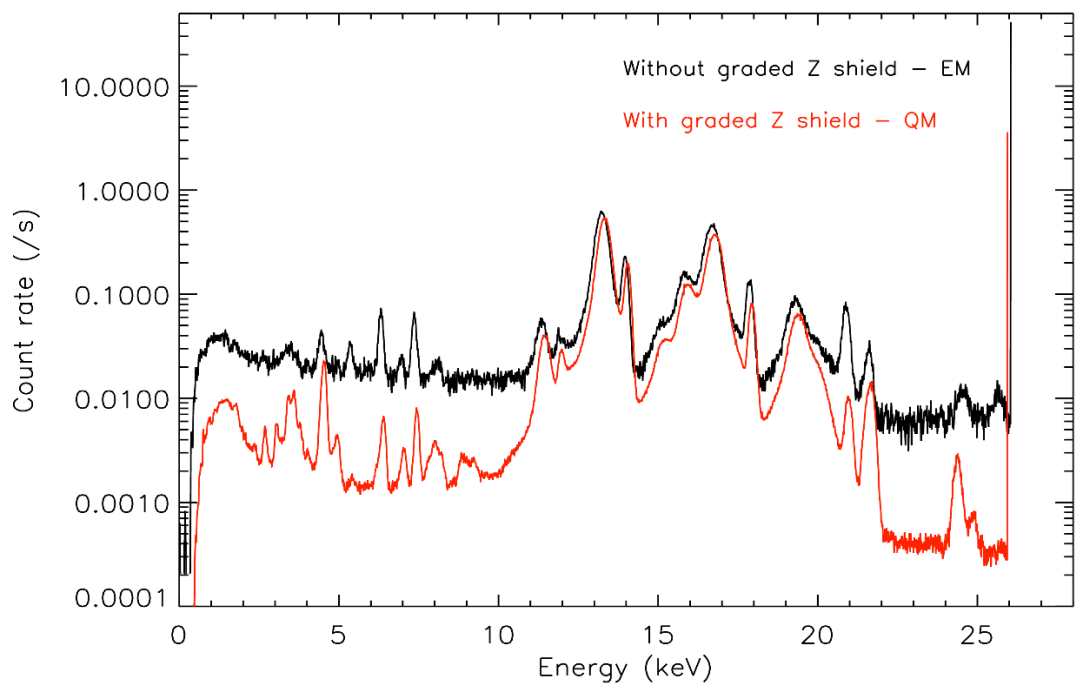

**Figure 13: Spectra with Perspex target for qualification and engineering models of APXS with and without multilayer shielding respectively demonstrating significant reduction in background counts with shielding.**

Flight model of the APXS is integrated with Chandrayaan-2 Pragyan rover and has undergone various integrated tests. Instrument performance was assessed at various stages using the same $^{55}$Fe calibration source and the performance is identical during all these tests. APXS payload is also characterized with the actual $^{244}$Cm sources during the rover thermo-vacuum test. The data was acquired from the calibration target plate when the APXS is in stowed condition and from an aluminium metallic plate in the deployed condition. Figure 14 shows the calibration plate spectrum with lines of all constituent elements identified. This spectrum will serve as a reference for in-flight verification of APXS performance. APXS flight model has undergone extensive ground calibration, both for the X-ray detector as well as the XRF and the PIXE processes using the flight radioactive sources, detailed results of which will be reported elsewhere.

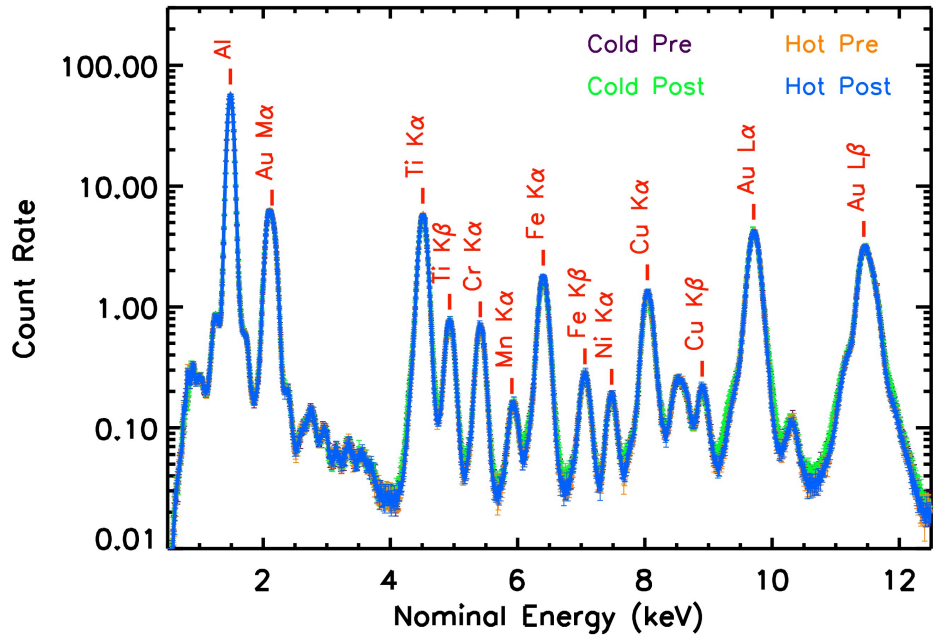

**Figure 14: Calibration target plate spectra acquired during integrated thermo-vacuum test with Chandrayaan-2 rover under hot and cold cycles.**

## In-flight operation plan and data analysis

Chandrayaan-2 lander is expected to land at dawn of a Lunar day. The lander and the rover operations are nominally planned till dusk, for a duration of approximately 14 Earth days. After the rover rolls out from the lander, a sequence of operations will be performed on the rover as planned. The rover stoppages on the lunar surface will be decided based on the feasibility of the rover movement, rover safety aspects and the location of scientific importance. The measurements with APXS are nominally planned at each rover stoppages. At each target location, calibration target plate observation is planned for ~5 minutes before and after the lunar target observation. After the pre-calibration data acquisition, APXS will be deployed and the fluorescence spectra from the lunar surface will be recorded for about 40 minutes. After completion, the APXS will be stowed back for post-calibration observation. Basic data from APXS is ADC value of the X-ray events detected. Data downloaded from Rover through Lander to Indian Space Science Data Center (ISSDC) undergoes level-0 processing. Higher level of processing generates gain corrected energy spectrum observed with APXS in FITS format compatible with standard X-ray spectral fitting tools. The data will be available publicly for download from ISSDC after a specified lock-in period.

## Summary

The flight model of the APXS instrument has been developed and characterized for performance requirement and also tested for various environmental conditions. APXS payload meets all the design criteria and provides the energy resolution of ~ 137 eV at 5.9 keV with low energy threshold of ~ 800 eV. APXS payload has been integrated with Chandrayaan-2 Pragyan rover and has gone through various environmental tests. Performance of the instrument is normal during all these tests. The APXS instrument will provide the first ever in-situ elemental composition measurements in the polar region of the Moon.

## Acknowledgements

APXS experiment is designed and developed by Physical Research Laboratory (PRL) Ahmedabad, supported by Department of Space. PRL is also responsible for the development of data processing software, overall payload operations and data analysis of APXS. The steering mechanism for APXS is provided by U. R. Rao Satellite Centre (URSC), Bengaluru. Thermal design and analysis of APXS was carried out by URSC, Bengaluru, whereas Space Application Centre (SAC) provided support in the mechanical design and analysis of APXS. SAC also supported in the fabrication of the flight model of the payload and its test & evaluation. Authors would like to thank various facilities and the technical teams of all above centers for their support during the design, fabrication and testing of this instrument. Authors also would like to thank the Indian Embassy in Moscow for their support for timely procurement of $^{244}$Cm sources. The Chandrayaan-2 project is funded by the Indian Space Research Organization (ISRO).

## References

1. McSween, H. Y., Huss, G. R., Cosmochemistry, Cambridge University Press, 2010.
2. Laxmiprasad, A. S., et al., An in situ laser induced breakdown spectroscope (LIBS) for Chandrayaan-2 rover: Ablation kinetics and emissivity estimates, Advances in Space Research, 52, 332-341, 2013.
3. Rieder, R., et al., Determination of the chemical composition of Martian soil and rocks: The alpha proton X ray spectrometer. J. Geophys. Res., 102, 4027-4044, 1997.
4. Rieder, R., et al., The new Athena alpha particle X-ray spectrometer for the Mars Exploration Rovers, Journal of Geophysical Research (Planets), 108, 8066, 2003.
5. Gellert, R., et al., The alpha-particle x-ray spectrometer (APXS) for the mars science laboratory (MSL) rover mission, Lunar and Planetary Science Conference, abstract no. 2364, 2009.